\documentclass[aps,pra,reprint,twocolumn,superscriptaddress,showpacs,showkeys,superscriptaddress,longbibliography]{revtex4-1}
\usepackage{amsmath,amssymb,amstext}
\usepackage[usenames,dvipsnames]{color}
\usepackage{graphicx}
\usepackage{bm,bbold,braket}
\usepackage{natbib}
\usepackage{txfonts, comment,stmaryrd}
\usepackage{dcolumn}
\usepackage{color}
\usepackage{xcolor}
\colorlet{rn}{red}
\colorlet{an}{blue}
\usepackage[english]{babel}

\usepackage[colorlinks,bookmarks=false,citecolor=blue,linkcolor=red,urlcolor=blue]{hyperref}
\begin{document}

\title{Droplet Arrays in Doubly-Dipolar Bose-Einstein condensates}
      \author{Ratheejit Ghosh}
      \affiliation{Department of Physics, Indian Institute of Science Education and Research Pune, Pune 411 008, India}   
      \author{Chinmayee Mishra}
      \affiliation{Department of Physics, Indian Institute of Science Education and Research Pune, Pune 411 008, India}   
      \affiliation{Indian Institute of Technology Gandhinagar, Gandhinagar 382 355, India}
      \author{Luis Santos}
      \affiliation{Institut f\"ur Theoretische Physik, Leibniz Universit\"at Hannover, Appelstrasse 2, DE-30167 Hannover, Germany}                   
\author{Rejish Nath}
       \affiliation{Department of Physics, Indian Institute of Science Education and Research Pune, Pune 411 008, India}
\date{\today}

\begin{abstract}
Gases of doubly-dipolar particles, with both magnetic and electric dipole moments, offer intriguing novel possibilities. We show that the interplay between doubly-dipolar interactions, quantum stabilization, and external confinement results in a rich ground-state physics of supersolids and incoherent droplet arrays in doubly-dipolar condensates. Our study reveals novel possibilities for engineering quantum droplets and droplet supersolids, including supersolid-supersolid transitions and the realization of supersolid arrays of pancake droplets. 
\end{abstract}

\maketitle

\section{Introduction}

The anisotropic and long-range nature of the dipole-dipole interactions leads to a rich physics in dipolar quantum gases, qualitatively different than that of their non-dipolar counterparts~\cite{bar08,bar12,lah09}, including anisotropic superfluidity~\cite{tic11,bis12,wen18}, roton-like excitations~\cite{cho18, pet18}, and the recent realization of quantum droplets~\cite{Kad16, fer16,cho16,sch16}. The latter result from the interplay between contact and dipolar interactions and the stabilization provided by quantum fluctuations~\cite{pet15}. Interestingly, the external confinement may result in the formation of arrays of droplets, which under proper conditions may remain mutually coherent, building a dipolar 
supersolid~\cite{lta18, bot19, cho19, nor21}, whose properties have recently been the focus of major attention \cite{nat19, guo19, zha19, tan21,ilz21,her21, zha21, pol21, nor21, bla22}.


Experiments on dipolar Bose-Einstein condensates have been realized so far with atoms with large permanent magnetic moments such as chromium~\cite{gri05, bea08}, erbium~\cite{aik12}, and dysprosium~(Dy)~\cite{min11}. Interestingly, a pair of quasi-degenerate states with opposite parity offers the possibility of inducing an additional electric dipole moment in Dy atoms using an electric field~\cite{lep18}. Recently, doubly-dipolar atoms and molecules possessing both electric and magnetic dipole moments~\cite{lep18, pio10,tom14,ree17,rva17, pas13,bar14,khr14,gut18} have attracted a large deal of interest due to their potential applications in quantum simulation~\cite{mic06}, computing~\cite{kar16}, tests of fundamental symmetries~\cite{hud11}, and for the tuning of collisions and chemical reactions~\cite{abr07}. Interestingly, the electric and magnetic moments may be oriented in different directions, opening novel possibilities for doubly-dipolar condensates~\cite{mis20}. Self-bound quantum droplets may undergo a dimensional crossover when varying the angle between the dipole moments without modifying the external confinement. 


In this paper, we show that the control of the relative angle between the 
two dipole moments opens new intriguing scenarios for quantum droplet arrays in 
doubly-dipolar condensates, including a density-modulated single droplet 
ground-state, supersolid-supersolid transitions, and the possibility of realizing an array of pancake-shaped quantum droplets.

%
The paper is structured as follows. In Sec.~\ref{dms}, we review a particular realization of a doubly-dipolar system using dysprosium atoms, 
employed in the rest of the paper. In Sec.~\ref{ddp}, we discuss the anisotropic properties of the doubly dipolar potential. In Sec.~\ref{hbec}, we introduce the extended Gross-Pitaevskii equation for a doubly-dipolar condensate, incorporating beyond-mean-field corrections. The properties of a single self-bound droplet are briefly discussed in Sec.~\ref{sbqd}. Section~\ref{tdc} is devoted to analyzing quantum droplet arrays in doubly-dipolar condensates. Finally, we summarize our conclusions in Sec.~\ref{summ}.


\section{Doubly-dipolar dysprosium atoms}
\label{dms}

In this section, we discuss a particular realization of a doubly-dipolar system using Dy atoms, briefly reviewing the proposal of Ref.~\cite{lep18}. However, other realizations, e.g. using molecules, should result in a similar physics.

In addition to its permanent magnetic moment, an electric moment may be induced in Dy atoms by an external electric field owing to a pair of quasi-degenerate states with opposite parity. These states, $|a\rangle$~(odd parity) and $|b\rangle$~(even parity), have total angular momenta $\{J_a=10, J_b=9\}$, and energies $\{E_a=17513.33 \  {\rm cm}^{-1}, E_b=17514.50 \ {\rm cm}^{-1}\}$. Within the electric-dipole approximation, the line-widths of the states are $\Gamma_a\approx  0$~(metastable) and $\Gamma_b=2.98 \times 10^4$ s$^{-1}$, respectively. We assume that the Dy atoms are in uniform magnetic and electric fields. The magnetic field, $\bm{B}=\mathcal B \hat{\bm{z}}$, is directed along $z$, setting the quantization axis and splitting the degeneracy of the energy levels  $E_a$ and $E_b$. The electric field, $\bm{E}=\mathcal E \hat{\bm{u}}$ mixes the Zeeman sublevels of the states, $\{|M_a=-J_a\rangle, ... , |+J_a\rangle, |M_b=-J_b\rangle, ... , |+J_b\rangle\}$, inducing an electric dipole moment along $\hat{\bm{u}}$. We assume that $\hat{\bm{u}}$ lies on the $xz$ plane forming an angle $\alpha$ with the $z$-axis. This relative angle plays a crucial role in the physics discussed below.



\begin{figure}[t!]
    \centering
  \includegraphics[width=0.8\columnwidth]{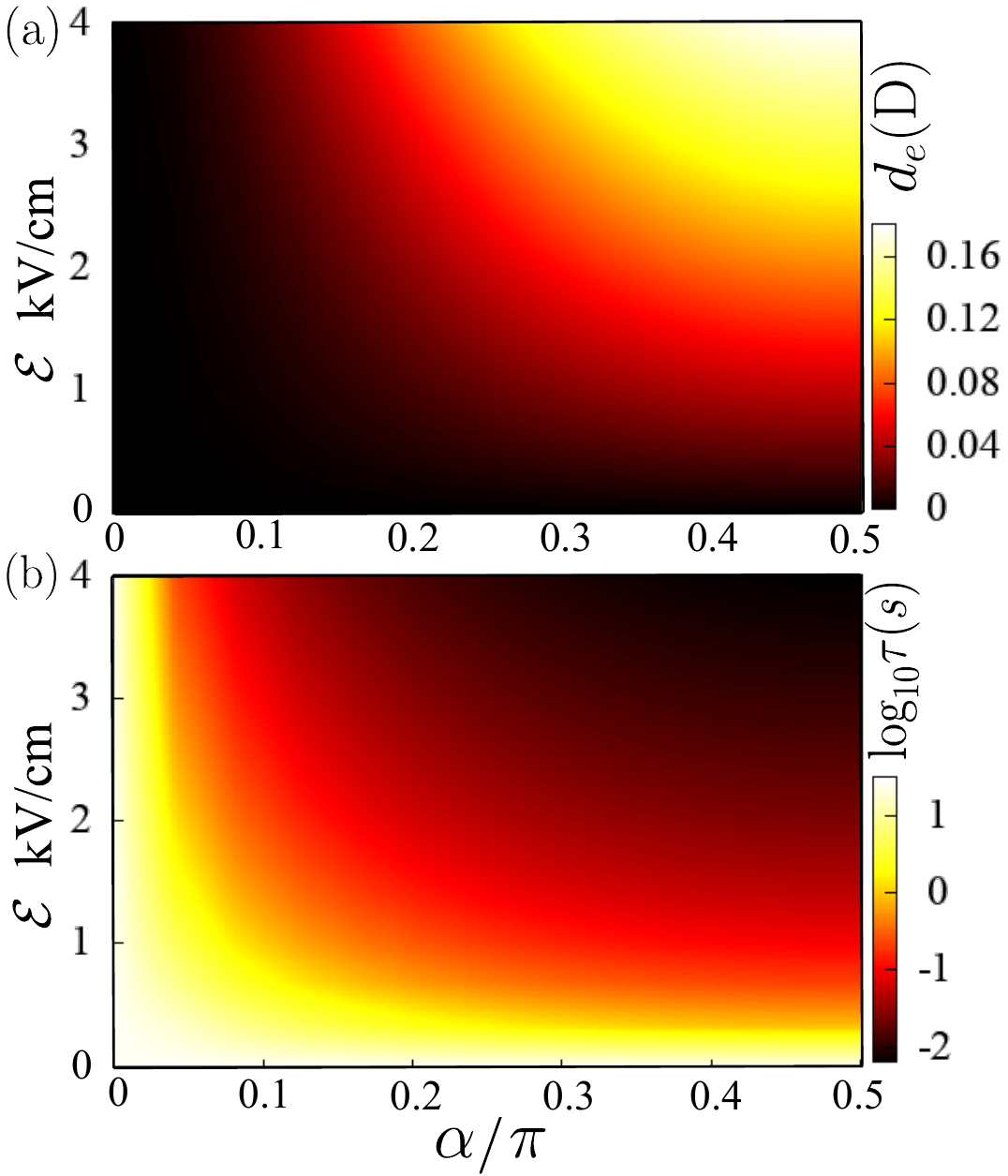}
    \caption{(Color online) (a) Electric dipole moment $d_e$ and (b) lifetime $\tau$ of the $|S\rangle$ state of a Dy atom, as a function of the electric field strength $\mathcal{E}$ and the angle $\alpha$ between the electric and magnetic fields for $\mathcal{B}=100$ G.} 
    \label{fig:1}
\end{figure}

Restricting to the subspace of both $E_a$ and $E_b$, the Hamiltonian for a Dy atom is $\hat H=\hat H_B+\hat H_{\rm{stark}}$ with
\begin{equation}
\hat{H}_B=E_a\sum_{M_a}|M_a\rangle\langle M_a|+E_b\sum_{M_b}|M_b\rangle\langle M_b|+\mu_B B (g_a M_a+g_b M_b), 
\end{equation}
where $g_a=1.3$ and $g_b=1.32$ are the Land\'e $g$ factors. The term $\hat{H}_{\rm{stark}}$ accounts for the interaction of the electric field with the Dy atom. The electric field strength is such that the lowest eigenstate of the atom is $|S\rangle= c_0|M_a=-10\rangle+\sum_i' c_i|i\rangle$ with $\sum_i'|c_i|^2/|c_0|^2\ll 1$, where the sum $\sum_i'$ is taken over all the magnetic sublevels except $|M_a=-10\rangle$, and $c_i$ is the probability amplitude for finding the atom in the state $|i\rangle$. The summation $\sum_i'$ has two contributions, one from the sublevels of $|a\rangle$ and the other from those of $|b\rangle$. Since $\Gamma_a\approx  0$, only the contributions from the sublevels $\{|M_b\rangle\}$ determine the lifetime of the stretched state $|S\rangle$, $\tau=(n_b\Gamma_b)^{-1}$, where $n_b$ is the total population in $\{|M_b\rangle\}$ sublevels. The magnetic and electric dipole moments of a Dy atom in $|S\rangle$ are, respectively:
\begin{eqnarray}
\label{dm}
d_m&=&-\mu_B\left(g_a\sum_{M_a=-J_a}^{J_a}|c_{M_a}|^2M_a+g_b\sum_{M_b=-J_b}^{J_b}|c_{M_b}|^2M_b\right)\\
d_e&=&-\dfrac{1}{\mathcal{E}}\sum_{M_a,M_b}c^*_{M_a}c_{M_b}\langle M_a|\hat{H}_{\rm{stark}}|M_b\rangle+\rm{c.c.}
\label{de}
\end{eqnarray}
with
\begin{eqnarray}
\langle M_a|\hat{H}_{\rm{stark}}|M_b\rangle&=&-\sqrt{\dfrac{4\pi}{3(2J_a+1)}}\langle a||\hat{d}||b\rangle \mathcal{E}\times  \nonumber \\
&&Y^*_{1,M_a-M_b}(\alpha,0)\, C^{J_aM_a}_{J_bM_b,1,M_a-M_b},
\end{eqnarray}
where, $\langle a||\hat{d}||b\rangle=8.16$ Debye is the reduced transition dipole moment, $Y_{l,m}(\theta,\phi)$ are the spherical harmonics and $C^{J_aM_a}_{J_bM_b,1,M_a-M_b}$ are the Clebsch-Gordan coefficients.

Figure~\ref{fig:1} depicts, for $\mathcal{B}=100$ G, $d_e$ and $d_m$ for the state $|S\rangle$, as a function of $\mathcal{E}$ and  $\alpha$. When $\alpha=0$, the spherical harmonics, $Y^*_{1,M_a-M_b}(\alpha,0)$ are non-zero only when $M_a=M_b$ and thus, the electric field couples pairs of sublevels with $M_a=M_b$. Since the state $|M_a=-10\rangle$ has no counterpart in the $\{|M_b\rangle\}$ subspace, the former is unaffected by the electric field. Hence, the electric dipole moment of the Dy atom in the state $|S\rangle$ vanishes for $\alpha=0$. When $\alpha$ grows, the electric field couples $|M_a=-10\rangle$ with other $|M_b\rangle$ sublevels, reaching a maximum mixing for $\alpha=\pi/2$~(see Fig.~\ref{fig:1}(a)). Therefore, for a given $\mathcal{E}$, $d_e$ increases with $\alpha$ until $\alpha=\pi/2$. This comes at the cost of decreasing the lifetime of $|S\rangle$, as shown in Fig.~\ref{fig:1}(b). For a range of experimentally realistic $\mathcal{E}=0$-$4$ kV/cm, $d_e$ varies from $0$ to $0.16$ Debye and the magnetic moment remains constant,  
$d_m \simeq 13\, \mu_B$~(higher than ground-state Dy atoms), whereas the lifetime of $|S\rangle$ varies from $28$ s~(considering electric-quadrupole and magnetic-dipole transitions) to $10$ ms~\cite{lep18}. 

\section{Doubly dipolar potential}
\label{ddp}
\begin{figure}
    \centering
    \includegraphics[width=0.7\columnwidth]{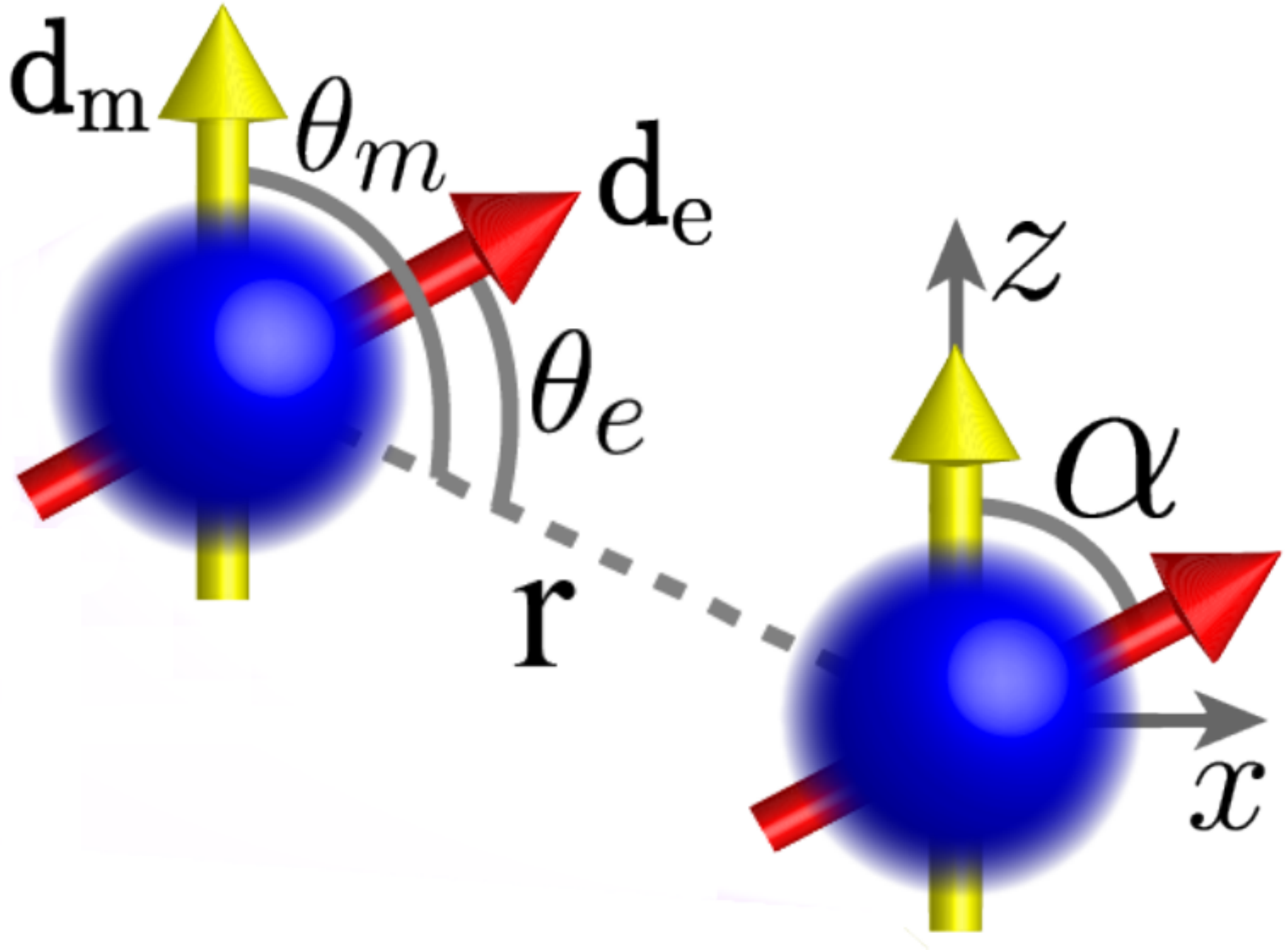}
 \caption{(Color online) Doubly-dipolar atoms. Both electric ($\bf{d}_e$) and magnetic ($\bf{d}_m$) dipoles are assumed polarized on the $xz$ plane, forming an angle $\alpha$ between them. The angle $\theta_m$~($\theta_e$) is the angle between $\bf{d}_m$~($\bf{d}_e$) and the vector joining the atoms, $\bf{r}$.}
    \label{fig:2}
\end{figure}

\begin{figure}
    \centering
    \includegraphics[width=\columnwidth]{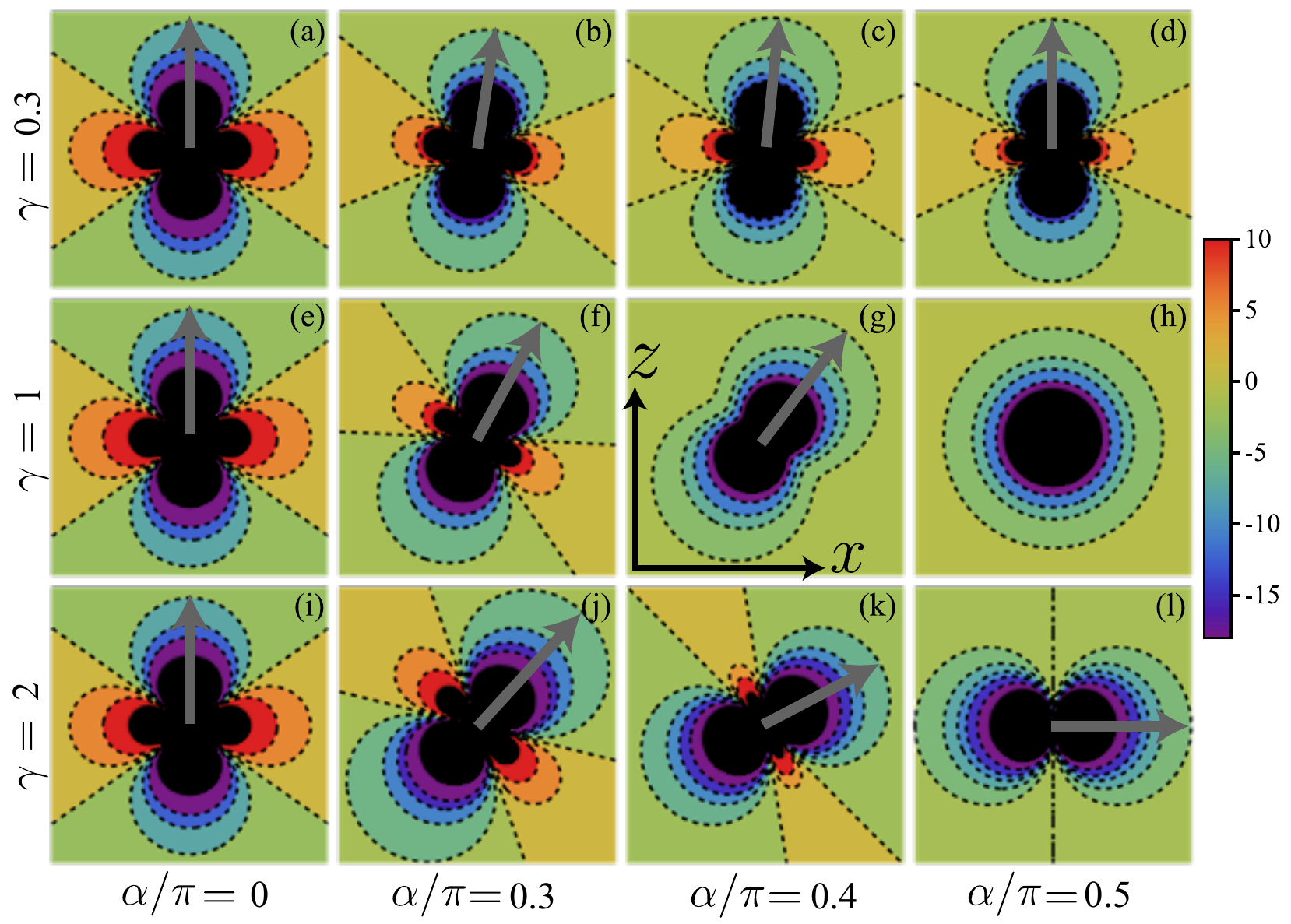}
   \caption{(Color online) Anisotropy of the doubly-dipolar potential on the $xz$-plane~($V_d^{y=0}(r,\theta)$) for different values of $\alpha$ and $\gamma$. Grey arrows indicate the effective polarisation axis determined by the polarization angle $\theta_p$~(Eq.~(\ref{tp1})).} 
    \label{fig:3}
\end{figure}

The doubly-dipolar interaction between two atoms is
\begin{equation}
V_d(\bm r)=\frac{\mu_0d_m^2}{4\pi}\frac{(1-3\cos^2\theta_m)}{|\bm{r}|^3}+\frac{d_e^2}{4\pi\epsilon_0}\frac{(1-3\cos^2\theta_e)}{|\bm{r}|^3}
\label{dddi}
\end{equation}
where $\mu_0$~($\epsilon_0$) is the vacuum permeability~(permittivity) and $\theta_m$~($\theta_e$) is the angle formed by the magnetic~(electric) dipole moment with the vector $\bm{r}$ joining the atoms~(Fig.~\ref{fig:2}). Whereas $V_d(\bm r)$ is always repulsive along the $y$-axis, it is anisotropic on the $xz$-plane. This anisotropy is well characterized by
the angular part of the dipolar potential on the $xz$-plane: 
\begin{eqnarray}
V_d^{y=0}(r,\theta)\propto\left[1-3\dfrac{\cos^2\theta+\gamma(\cos\theta\cos\alpha+\sin\alpha\sin\theta)^2}{1+\gamma}\right],
\label{anixz}
\end{eqnarray}
where $\theta$ is the polar angle, and $\gamma=(d_e/d_m)^2/(\mu_0\epsilon_0)$ characterizes the relative strength between the electric and magnetic dipole moments. The ratio $\gamma$ can be varied independently of $\alpha$ by tuning $\mathcal E$. In Fig.~\ref{fig:3}, we depict $V_d^{y=0}({\bm r})$ for different $\alpha$ and $\gamma$. When $\alpha=0$, we have the usual dipolar potential, attractive along $z$ and repulsive along $x$~\cite{lah09}. As $\alpha$ increases up to $\pi/2$, 
the dependence of the potential on $\gamma$ becomes more significant. For $\alpha=\pi/2$, when $\gamma$ grows the potential inverts eventually its anisotropy (last column of Fig.~\ref{fig:3}). If $-4\gamma^2+\gamma-4 \leq 9\gamma$, there exists a critical angle $\frac{1}{2}\cos^{-1}\left[(-4\gamma^2+\gamma-4)/9\gamma\right]$ above which the $xz$ potential becomes purely attractive~(see Figs.~\ref{fig:3}~(g),~(h), and~(l)), but remains anisotropic except when $\alpha=\pi/2$ and $\gamma=1$. For that case  $V_d^{y=0}(r,\theta)=-1/r^3$.



\begin{figure}
    \centering
    \includegraphics[width=0.9\columnwidth]{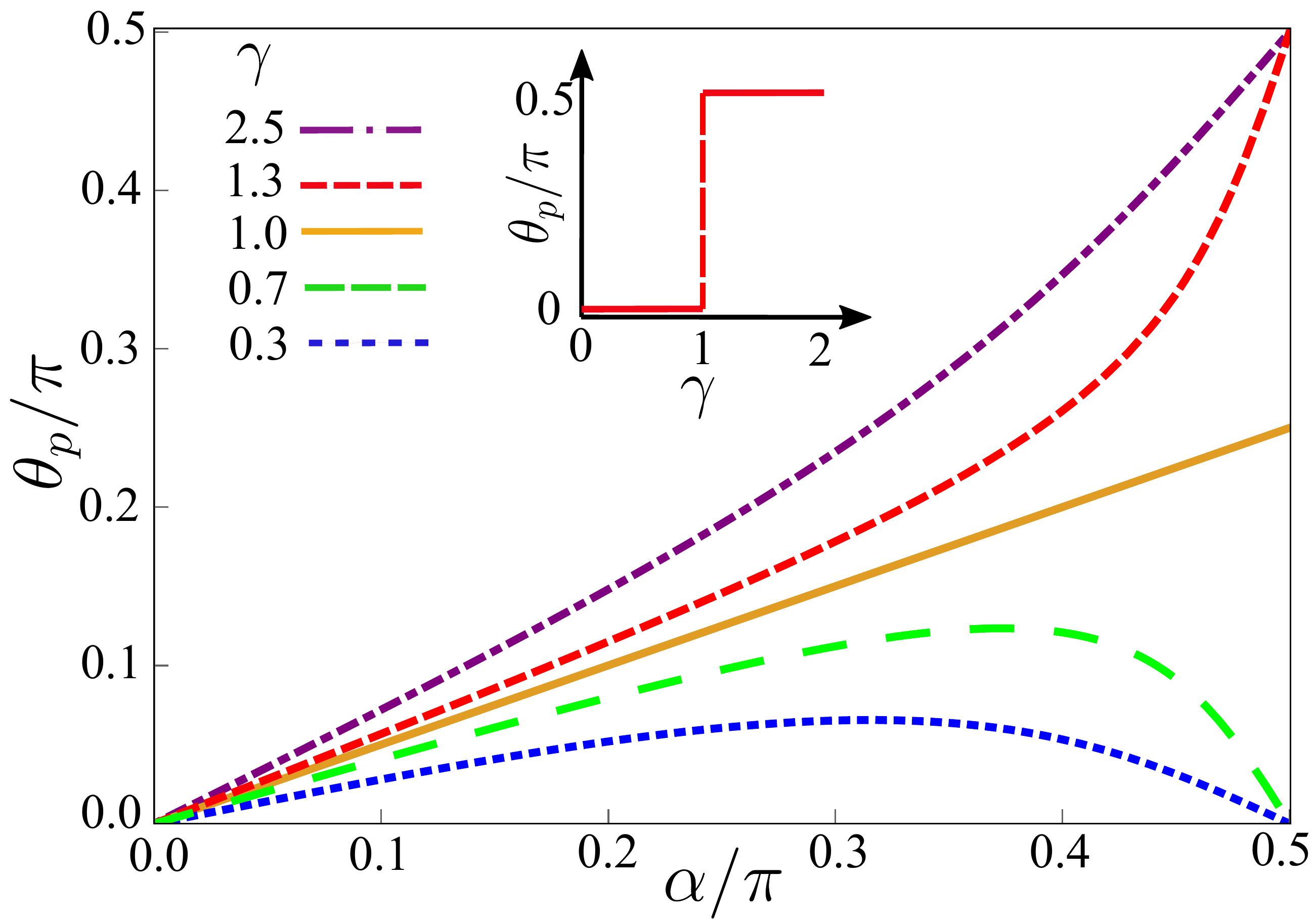}
    \caption{(Color online) Polarization angle $\theta_p$ as a function of $\alpha$ for different values of $\gamma$. The inset shows $\theta_p$ as a function of $\gamma$ for $\alpha=\pi/2$, exhibiting a jump at $\gamma=1$.} 
    \label{fig:4}
\end{figure}

Despite this nontrivial anisotropy, we may define an effective polarization axis, depicted by arrows in Fig.~\ref{fig:3}, given by the direction in which the potential is maximally attractive. This direction lies on the $xz$ plane, sustaining an angle
\begin{equation}
\theta_p(\alpha,\gamma)=\cos^{-1}\left[\frac{1}{\sqrt{2}}\sqrt{1+\frac{1+\gamma\cos2\alpha}{\sqrt{1+\gamma^2+2\gamma\cos2\alpha}}}\right],
\label{tp1}
\end{equation}
with the positive $z$-axis. As shown in Fig.~\ref{fig:4}, 
for a dominant magnetic dipole~($\gamma<1$), 
$\theta_p$ increases with $\alpha$, reaches a maximum  [$\theta_p^{max}=\cos^{-1}\left(\frac{1}{2}[1+\sqrt{1-\gamma^2}]\right)^{1/2}$] at $\alpha=\frac{1}{2}\cos^{-1}(-\gamma)$ and then decreases back to zero at $\alpha=\pi/2$. 
On the contrary, for a dominant electric dipole~($\gamma>1$), $\theta_p$ increases monotonously from zero to $\pi/2$. A linear relation, $\theta_p=\alpha/2$ holds for $\gamma=1$. When $\alpha=\pi/2$ and $\gamma=1$, $\theta_p$ is not defined due to the isotropic nature of the $xz$-interactions. Thus, $\theta_p$ exhibits a discontinuous behavior as a function of $\gamma$ for $\alpha=\pi/2$, changing abruptly from zero to $\pi/2$ across $\gamma=1$~(inset of Fig.~\ref{fig:4}). As discussed below, $\theta_p$ plays a key role in determining the properties of doubly-dipolar droplets.



\section{Extended Gross-Pitaevskii equation}
\label{hbec}
At this point, we consider a condensate of $N$ doubly-dipolar Dy bosonic atoms of mass $M$. The condensate wavefunction $\psi(\bm r,t)$ is given in mean-field theory by the nonlocal Gross-Pitaevskii equation:  $i\hbar\dot{\psi}(\bm r,t)=\mathcal H\psi(\bm r,t)$, with 
\begin{eqnarray}
\mathcal H=\dfrac{-\hbar^2\nabla^2}{2M}+V_{ext}(r)+\int d^3r'V(\bm r-\bm r')|\psi(\bm r', t)|^2,
\label{nlgpe}
\end{eqnarray}
where $V_{ext}(r)=M(\omega_x^2x^2+\omega_y^2y^2+\omega_z^2z^2)$ is the external harmonic confinement, $V(\bm r)=N \left (g\delta(\bm r)+V_d(\bm r) \right )$ is the interaction potential, including contact and doubly-dipolar interactions. The coupling constant $g=4\pi\hbar^2 a_s/M$ characterizes the contact interaction, with $a_s$ the $s$-wave scattering length. To quantify the strength of the dipolar interactions, we introduce the constants $g_m=N\mu_0d_m^2/4\pi$, $g_e=Nd_e^2/4\pi\epsilon_0$, and $\gamma=g_e/g_m$. 

For a homogeneous condensate~($V_{ext}(r)=0$) of density $n_0$, the Bogoliubov excitations are
\begin{eqnarray}
\varepsilon_{\bm k} =\sqrt{\dfrac{\hbar^2 k^2}{2M}\left(\dfrac{\hbar^2 k^2}{2M}+2g_m n_0\left[\beta+\mathcal F(\theta_k,\phi_k,\alpha)\right]\right)}
\label{disp3d}
\end{eqnarray}
where $k$ is the quasi-momentum, $\beta=g/g_m$, and
\begin{eqnarray}
\mathcal F(\theta_k,\phi_k,\alpha)&=&\frac{4\pi\gamma}{3}\left[3\left(\cos\alpha\cos\theta_k+\sin\alpha\sin\theta_k\cos\phi_k\right)^2-1\right] \nonumber \\
&&+\frac{4\pi}{3}(3\cos^2\theta_k-1),
 \label{fd}
\end{eqnarray}
with $\theta_k$ and $\phi_k$ the angular coordinates in momentum space. The phonon modes $\varepsilon_{\bm k\to 0}=c(\theta_k,\phi_k)\hbar k$ determine the stability properties of the condensate, where 
\begin{equation}
c(\theta_k,\phi_k)=\left[g_m n_0\left(\beta+\mathcal F(\theta_k,\phi_k,\alpha)\right)/M\right]^{1/2},
\label{sv}
\end{equation} 
 is the direction-dependent sound velocity. The stiffest phonons~(largest $c$) propagate along the effective polarization axis set by $\theta_p$, whereas the softest ones are perpendicular to it.  For dipoles polarized on the $xz$ plane, phonons propagating along $y$ are always soft, and determine the stability criteria, i.e. $c_y^2=c^2(\pi/2,\pi/2)=c_m^2\left[\beta-\frac{4\pi}{3}(1+\gamma)\right]<0$ where $c_m=\sqrt{g_mn_0/M}$. Thus, a homogeneous doubly-dipolar BEC becomes unstable against local collapses if $\beta<\frac{4\pi}{3}(1+\gamma)$.  
Using the dispersion in Eq.~\eqref{disp3d}, we obtain the Lee-Huang-Yang~(LHY) correction to the ground state energy:
\begin{eqnarray}
\Delta E&=&\dfrac{V}{2}\int\dfrac{d^3q}{(2\pi)^3} \left[\varepsilon_q-\dfrac{\hbar^2q^2}{2m}-nV_{q}+\dfrac{mn^2V_{q}^2}{\hbar^2q^2}\right],\nonumber
\end{eqnarray}
where $V$ is the volume and $V_{q}$ is the Fourier transform of $V(\bm r)$. After integrating over $k$, we get the LHY correction to the chemical potential $\Delta\mu=\partial\Delta E/\partial N$ \cite{wac16, bis16,wac16-2,sai16}:
\begin{eqnarray}
\Delta\mu=\frac{g_m^{5/2}}{3\pi^3N}\left(\frac{Mn_0}{\hbar^2}\right)^{3/2}  \int d\Omega_k\left[\beta+\mathcal F(\theta_k,\phi_k,\alpha)\right]^{\frac{5}{2}},
\label{lhy}
\end{eqnarray}
where $\int d\Omega_k=\int_0^{2\pi}d\phi_k\int_0^\pi d\theta_k\sin\theta_k$. The correction, $\Delta\mu$ becomes complex when $\beta<\frac{4\pi}{3}(1+\gamma)$ for which the homogeneous doubly-dipolar BEC is unstable. The real part of $\Delta\mu$ is dominated by hard modes, whereas the unstable low-momentum excitations determine the imaginary part. Not very deep in the instability regime, ${\rm Im}[\Delta\mu]/{\rm Re}[\Delta\mu] \ll 1$ and ${\rm Im}[\Delta\mu]$ can be disregarded when analyzing the physics of doubly-dipolar condensates.  For a finite size condensate, ${\rm Im}[\Delta\mu]$ is further suppressed by a low-momentum cut-off \cite{bis16, sai16, wac16}.

The LHY correction $\Delta\mu$ is repulsive and has a density dependence of $n_0^{3/2}$. Because of this density dependence, the LHY correction becomes significant at high densities, stabilizing the condensate against mean-field collapse. Incorporating the LHY correction into the Gross-Pitaevskii equation in local density approximation~($n_0\to n(\bm r, t)$)~\cite{lim11, wac16, bis16,wac16-2,sai16, bai16, bai17}, we obtain the extended Gross-Pitaevskii equation~(eGPE):
 \begin{equation}
 i\hbar\dot{\psi}(\bm r,t)=\left(\mathcal H+\Delta\mu\left[n(\bm r, t)\right]\right)\psi(\bm r,t).
 \label{ggpe}
 \end{equation}
Below, we numerically solve Eq.~\eqref{ggpe} via imaginary time evolution to obtain the ground states of a doubly-dipolar BEC.
 


\begin{figure}
    \centering
    \includegraphics[width=1\columnwidth]{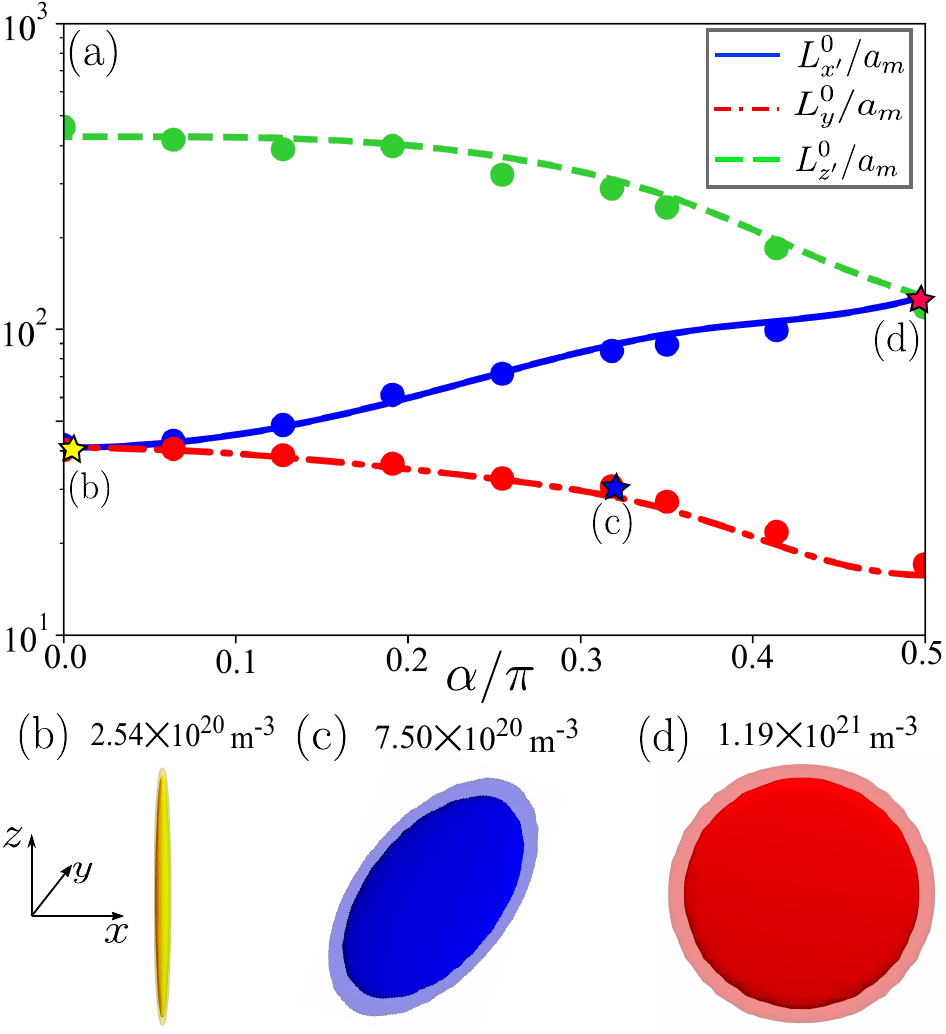}
     \caption{(Color online) (a) Equilibrium widths of a self-bound droplet as a function of $\alpha$ for $N=2$k, $\gamma=1$ and $a_s=200a_0$. The numerical results of Eq.~(\ref{ggpe})~(solid points) are in very good agreement with that of variational calculations~(solid and dashed lines). (b)-(d) show the density iso-surface of the ground states of the self-bound droplets obtained via imaginary time evolution of Eq.~\eqref{ggpe} at $\alpha/\pi=0$, $0.32$ and $0.5$, respectively. The peak density of the droplet is provided at the top for each case. We observe a structural transformation from a cigar to pancake shape as a function of $\alpha$.}  
    \label{fig:5}
\end{figure}


\section{Self-bound droplet} 
\label{sbqd}
Before discussing the properties of droplet arrays, it is convenient to briefly review the properties of individual doubly-dipolar quantum droplets~\cite{mis20}, 
which may be well understood using 
a variational Gaussian ansatz
\begin{eqnarray}
\psi({\bm r}, t)&=&\dfrac{1}{\pi^{3/4}\sqrt{L_x'L_yL_z'}}\exp\left[-\dfrac{x'^2}{2L_x'^2}-\dfrac{y^2}{2L_y^2}-\dfrac{z'^2}{2L_z'^2}+ \right.\nonumber \\
&&\left. ix'^2\beta_{x}+iy^2\beta_y+iz'^2\beta_{z}+ix'z'\beta_{xz}\right],
\label{vg0}
\end{eqnarray}
with $x'=x\cos\theta-z\sin\theta$, and $z'=x\sin\theta+z\cos\theta$. 
The variational parameters are the orientation angle $\theta$ of the droplet 
on the $xz$-plane, and $L_x', L_y$, and $L_z'$, the droplet widths along $x'$, $y$, and $z'$, respectively. 
The droplet minimizes its energy by orienting along the 
effective polarization direction~($\theta=\theta_p$).

Figure ~\ref{fig:5}(a) shows the equilibrium widths, $L_{x',y,z'}^0$, as a function of $\alpha$, for $\gamma=1$, $N=2000$, and $a_s=200 a_0$. Changing $\alpha$ results in a dimensional crossover. For $\alpha=0$, the droplet is cigar-shaped~($L_{x'}^0=L_y^0\ll L_{z'}^0$), see Fig.~\ref{fig:5}(b). As $\alpha$ increases, the effective polarization axis tilts away from the $z$-axis, and the repulsive interaction along the $x$-axis is reduced. The latter causes an increase of $L_{x'}^0$, and a decrease of $L_y^0$ and $L_{z'}^0$, giving a completely anisotropic droplet, as in Fig.~\ref{fig:5}(c). When $\alpha$ approaches $\pi/2$, the droplet acquires a pancake shape, reaching $L_{x'}^0=L_{z'}^0\gg L_{y}^0$ at $\alpha=\pi/2$~(see Fig.~\ref{fig:5}(d)). 
Hence, whereas droplets are cigar-like in usual dipolar condensates ~\cite{Kad16, fer16, cho16, sch16}, doubly-dipolar condensates open the interesting possibility of a controllable modification of quantum droplets from cigar- to pancake-shaped. 


\section{Doubly-dipolar droplet arrays}
\label{tdc}
The presence of an external confinement may result in multi-droplet ground states, as observed in condensates of magnetic atoms~\cite{bai18,bot19,nor21,pol21}. Under proper conditions, the droplets may keep mutual phase coherence, resulting in dipolar supersolids. In this section, we investigate the novel possibilities opened by the doubly-dipolar potential in the context of droplet arrays and supersolids.

%

\begin{figure}[t!]
    \centering
    \includegraphics[width=\columnwidth]{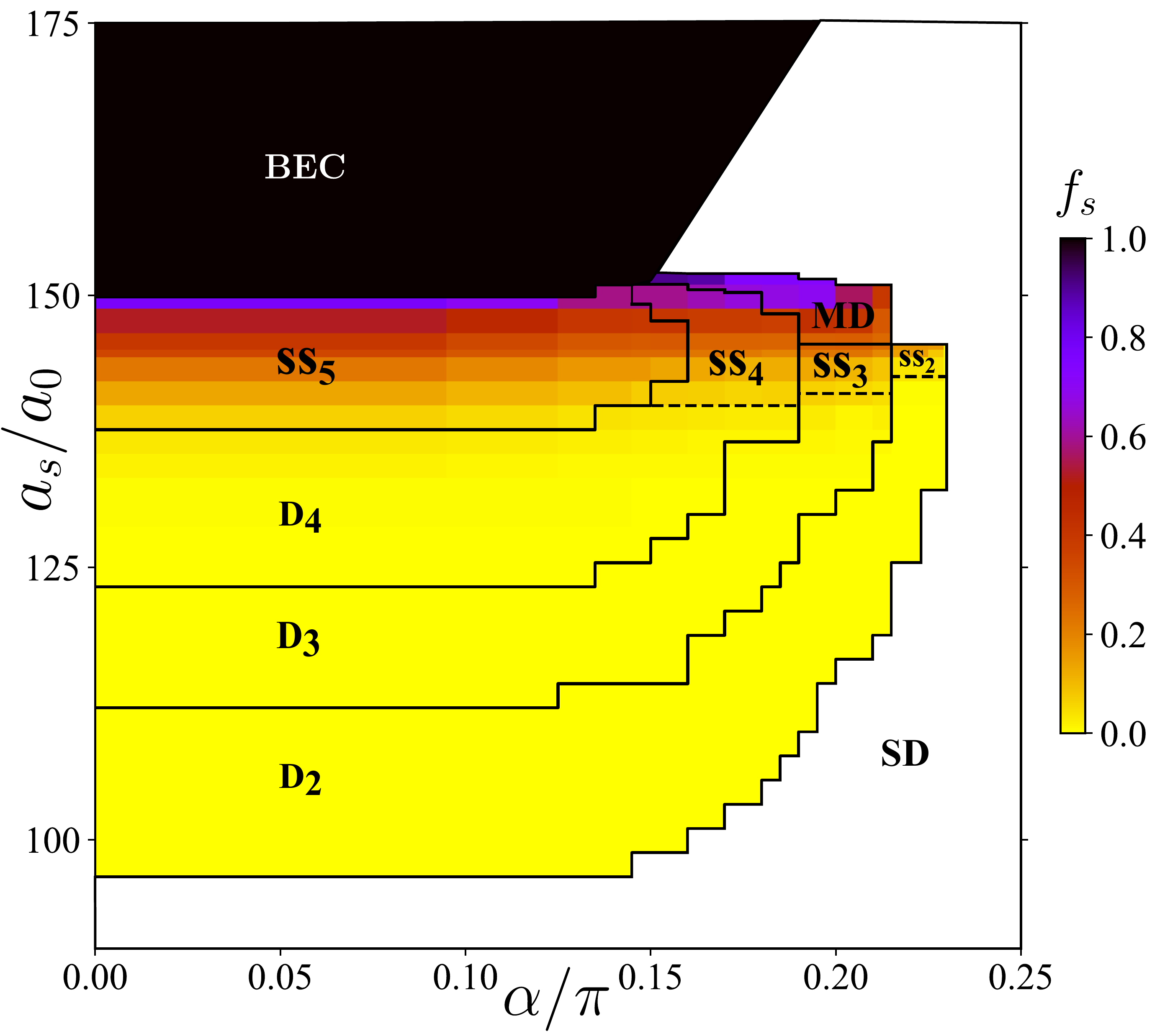}
   \caption{(Color online) Ground-state phase diagram as a function of $\alpha$ and $a_s$ for $\omega_{x,y,z}=2\pi \times (18.5,53,81)$ Hz, $N=35000$, $\mathcal B=100$ G and $\mathcal E=2.68$ kV/cm. The incoherent~(supersolid) arrays with $n$ droplets are denoted as ${\rm D_n}$~(${\rm SS_n}$). The color bar shows the superfluid fraction $f_s$ of the multi-droplet states. Solid lines separate states with different number of droplets whereas dashed lines separate supersolid and incoherent arrays with the same droplet number. MD is the region where we observe a density modulated droplet, and SD is the regime for a single droplet. To distinguish between BEC and SD, the superfluid fraction is not shown in the SD region.}    
 \label{fig:6}
\end{figure}


In the following, we consider a doubly-dipolar Dy condensate of $N=35000$ atoms in a magnetic field of $\mathcal B=100$ G and an electric field $\mathcal E=2.68$ kV/cm~(experimentally more convenient than the values considered in the previous section). For these field strengths, $\gamma=1$ for $\alpha=\pi/2$, and the lifetime of state $|S\rangle$ varies from $28$ s at $\alpha=0$ to $58$ ms at $\alpha=\pi/2$~(experimentally sufficient to observe the physics discussed here). As in the previous section, the effective polarization direction is assumed to lie on the $xz$ plane. To distinguish between incoherent droplets and supersolids, we employ Legget's upper bound of the superfluid fraction \cite{leg70}: 
\begin{equation}
    f_s = (2L)^2 \left[\int_{-L}^{L}dq \Tilde{n}(q) \int_{-L}^{L} \frac{dq}{\Tilde{n}(q)}\right]^{-1},
\end{equation}
where $q$ is the coordinate along which the droplet array is formed, and $\tilde{n}(q)$ is the column density obtained after integrating over the other two axes. The length $2L$ encloses the central region, 
where droplets form. For a mean-field stable condensate (BEC regime) and an unmodulated single droplet~(SD regime), $f_s=1$, whereas $f_s \sim 0$ for an incoherent droplet array (${\rm D_n}$) with $n>1$. Intermediate $f_s$ values characterize the supersolid regime. We employ the criterion $f_s>0.1$ to identify a regime as supersolid. In the following, we discuss separately the case of weaker confinement on the dipole plane and orthogonal to it.
\subsection{Weaker confinement on the dipole plane}
\label{ftg1}

Figure~\ref{fig:6} illustrates the possible ground-states as a function of $\alpha$ and $a_s$ for the case of weaker confinement on the dipole 
plane. We consider $N=35000$ atoms in a trap elongated along $x$, with $\omega_{x,y,z}=2\pi \times (18.5,53,81)$ Hz. 
In the diagram, incoherent~(supersolid) droplet arrays are labelled as ${\rm D_n}$~(${\rm SS_n}$), where $n$ stands for the number of droplets. For $\alpha=0$, we retrieve the known physics of usual dipolar condensates. When decreasing $a_s$, the ground-state transitions from an unmodulated~(denoted as BEC) regime to a supersolid and eventually to incoherent droplets~\cite{lta18, bot19, cho19}. For lower $a_s$, the dipolar interactions become more dominant, leading to fewer incoherent droplets, and eventually to a single one~(SD regime). 
The phase diagram remains unchanged for $\alpha\lesssim 0.12\pi$, since the electric dipole moment is very small~(see Fig.~\ref{fig:1}(a)). 
%


\begin{figure}
    \centering
    \includegraphics[width=1\columnwidth]{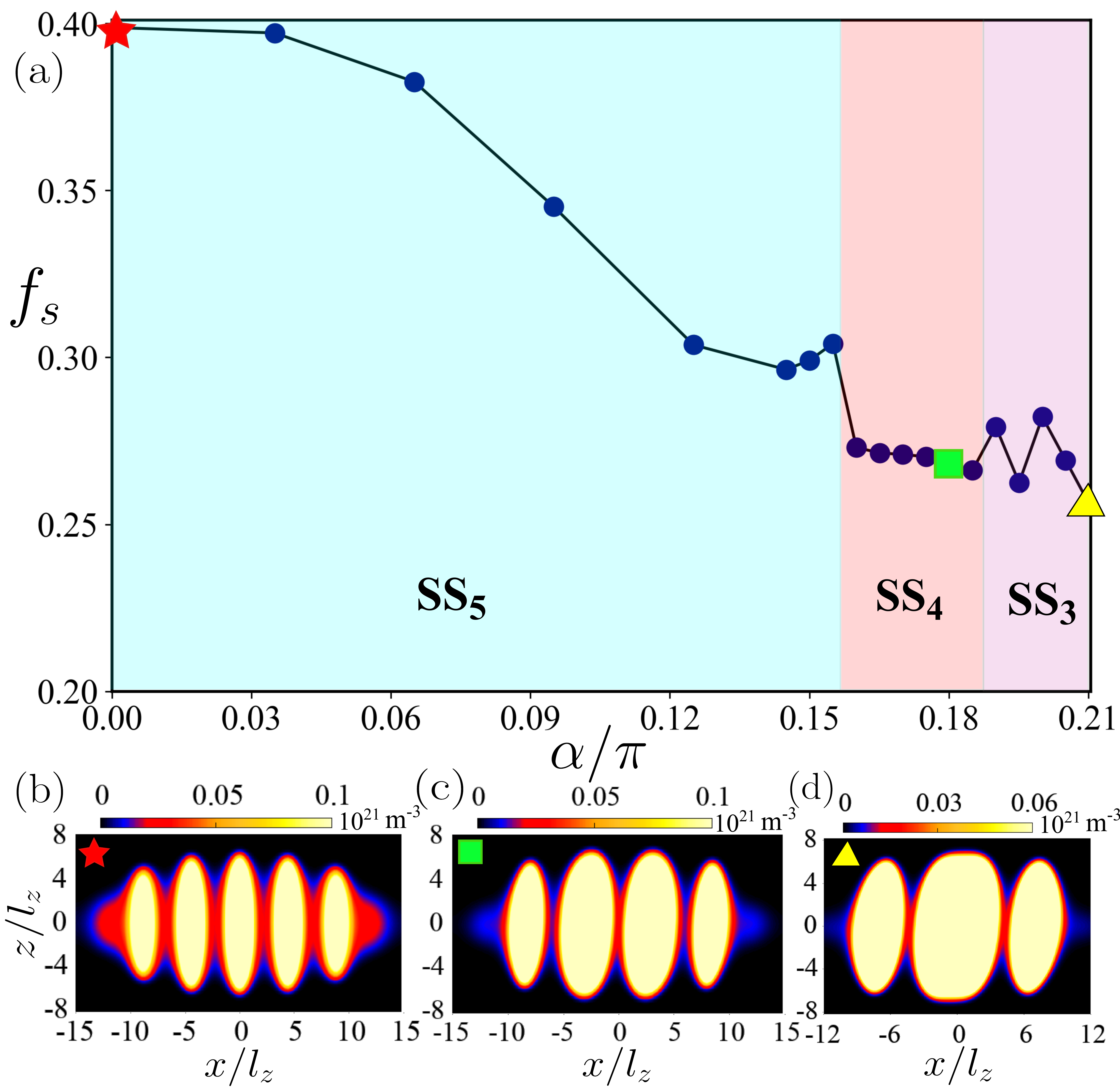}
    \caption{(Color online) Supersolid-supersolid transitions. Superfluid fraction $f_s$ as a function of $\alpha$ for $a_s=145a_0$. The other parameters are same as in Fig.~\ref{fig:6}. (b)-(d) show supersolid densities ($|\psi(x, y=0, z)|^2$) at $\alpha/\pi=0$~(${\rm SS}_5$), $0.18$~(${\rm SS}_4$), and $0.21$~(${\rm SS}_3$), respectively.}  
    \label{fig:7}
\end{figure}

In contrast, the phase diagram is radically altered for larger $\alpha$. Remarkably, the ground state may undergo, as a function of $\alpha$, a transition between supersolid phases with a different number of droplets. These transitions are performed while keeping a significant superfluid fraction~(see Fig.~\ref{fig:7}). Note as well that due to the changing anisotropy of the doubly-dipolar potential, varying $\alpha$ 
results in a modification of the shape and orientation of the droplets that form the supersolid~(see Figs.\ref{fig:7}~(b-d)). 


\begin{figure}[t!]
    \centering
    \includegraphics[width=1\columnwidth]{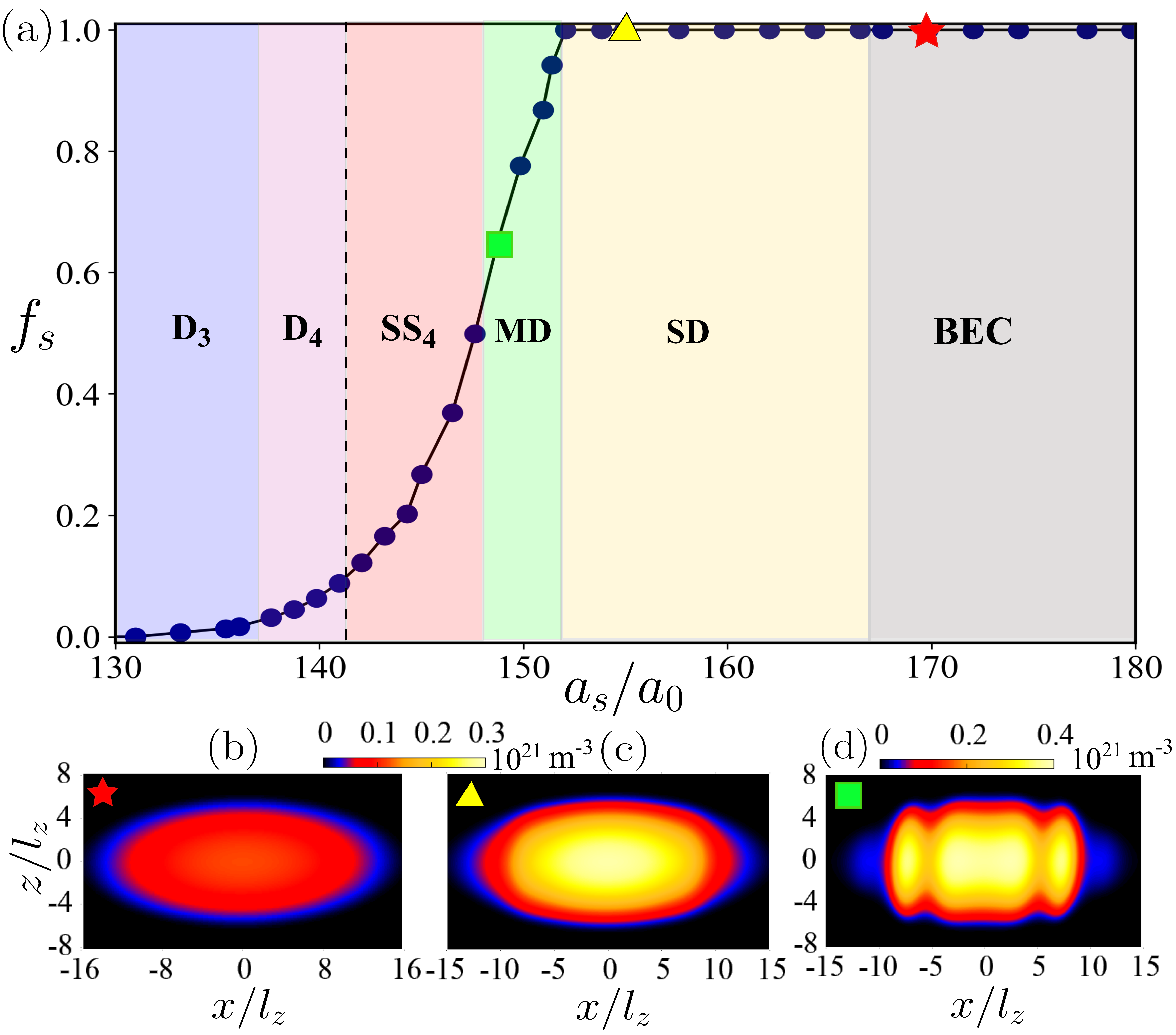}
    \caption{(Color online) Superfluid fraction $f_s$ as a function of $a_s$ for $\alpha/\pi=0.18$. The other parameters are the same as in Fig.~\ref{fig:6}. Dashed line separates supersolid and incoherent arrays with the same droplet number. (b)-(d) show the densities $|\psi(x, y=0, z)|^2$. Upon decreasing $a_s$, the BEC phase~(depicted in (b) for $a_s=170 a_0$)   transitions into a single droplet~(shown in (c) for $a_s=155 a_0$), which develops density modulations~(as seen in (d) for $a_s=149 a_0$).} 
    \label{fig:8}
\end{figure}
\begin{figure}
    \centering
    \includegraphics[width=1\columnwidth]{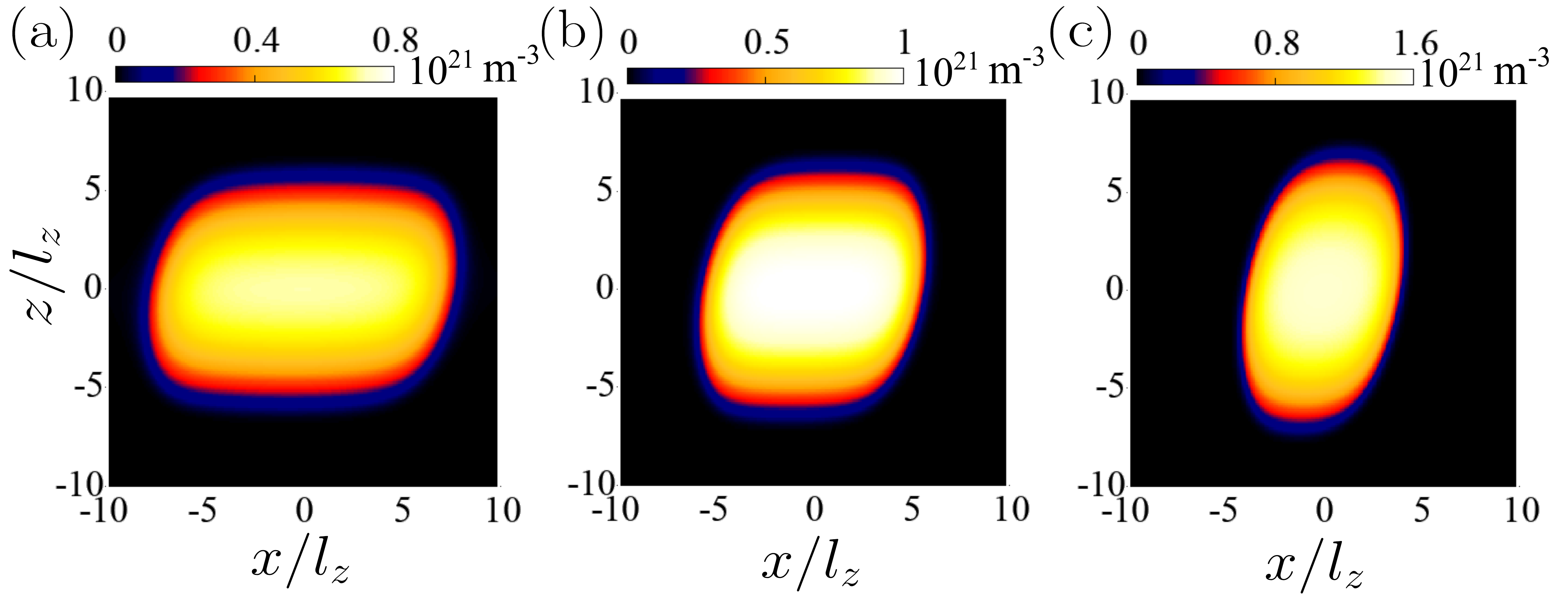}
    \caption{(Color online) Droplet density ($|\psi(x, y=0, z)|^2$) in the $xz$-plane for $\alpha/\pi=0.25$, and $a_s/a_0=155$~(a), $146$~(b), and $135$~(c). The other parameters are the same as in Fig.~\ref{fig:6}. The droplet acquires a rectangular cuboid shape with its aspect ratio controlled by $a_s$.} 
    \label{fig:9}
\end{figure}

 For a sufficiently large $\alpha\gtrsim 0.15\pi$, the mean-field stable condensate transitions for decreasing $a_s$ into a single droplet~(see Fig.~\ref{fig:8}), resembling the situation found in usual dipolar condensates for a small-enough particle number~\cite{cho16}. For $\alpha\gtrsim 0.23\pi$, the single droplet remains the ground-state when further decreasing $a_s$. However, the droplet shape may depart very significantly from the typical elongated form found in usual dipolar condensates. The interplay between the external confinement and the doubly-dipolar potential causes shearing and tilting, leading to rectangular cuboid shapes, with an aspect ratio controlled by $a_s$~(see Fig.~\ref{fig:9}). 

The situation is very different for $0.15\pi\lesssim\alpha\lesssim 0.23\pi$, where the single droplet acquires a density modulation for decreasing $a_s$ due to the roton-like softening of the lowest droplet mode along the $x$-axis~\cite{pal20}. When further decreasing $a_s$~(see Fig.~\ref{fig:8}), this modulated droplet~(MD) supersolid ground-state evolves into a droplet supersolid and then into an incoherent droplet array, with a decreasing number of droplets until reaching back a single-droplet solution.


\subsection{Weak confinement perpendicular to the dipole plane}
\label{ftg2}

\begin{figure}
    \centering
    \includegraphics[width=1\columnwidth]{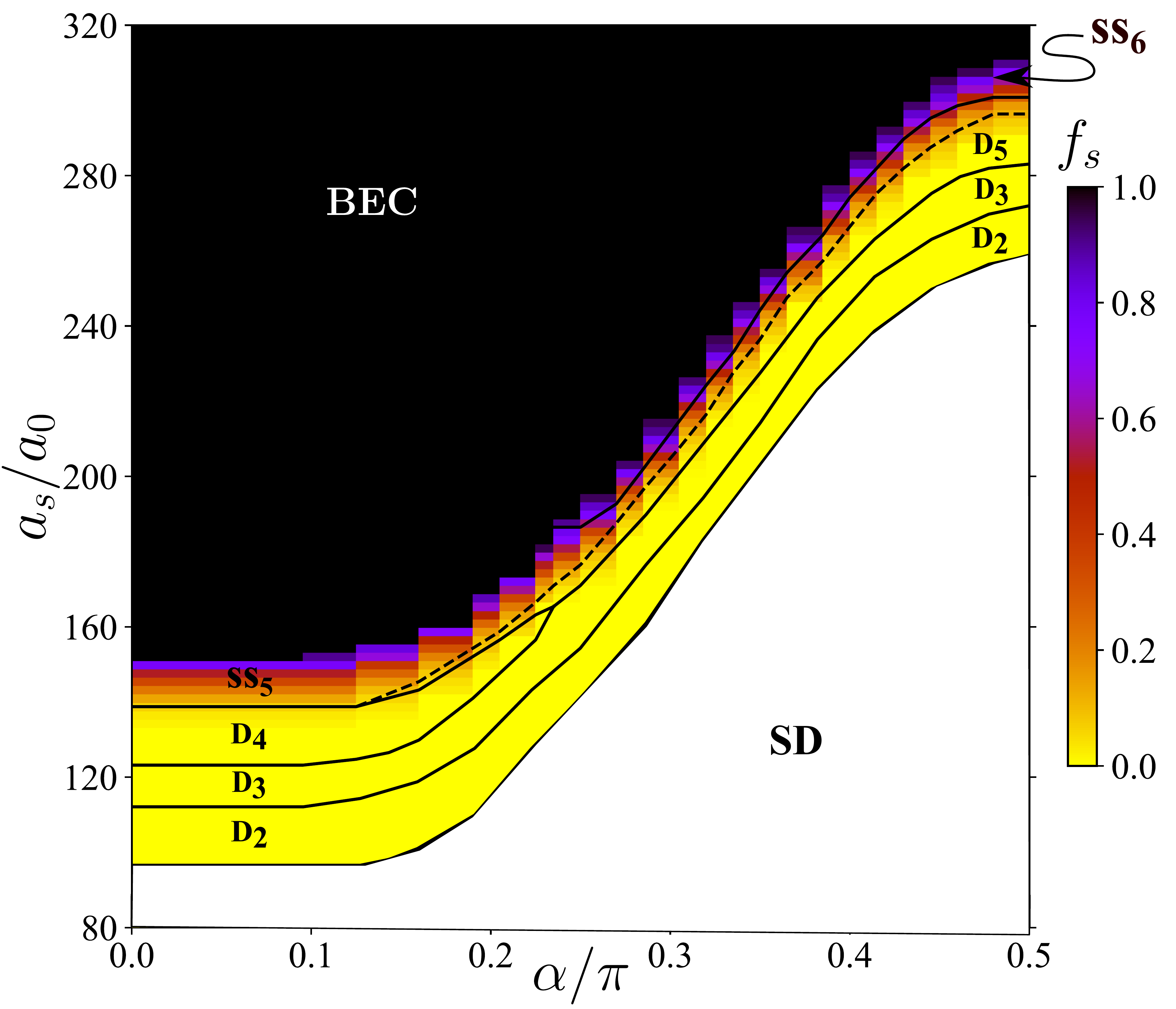}
   \caption{(Color online) Ground-state phase diagram as a function of $\alpha$ and $a_s$ for $\omega_{x,y,z}=2\pi \times (53, 18.5, 81)$ Hz. The other parameters are the same as in Fig.~\ref{fig:6}. The color bar shows the superfluid fraction $f_s$ of the multi-droplet states. Solid lines separate states with different number of droplets, whereas dashed lines separate supersolid and incoherent arrays with the same droplet number.}    \label{fig:10}
\end{figure}

\begin{figure}
    \centering
    \includegraphics[width=1\columnwidth]{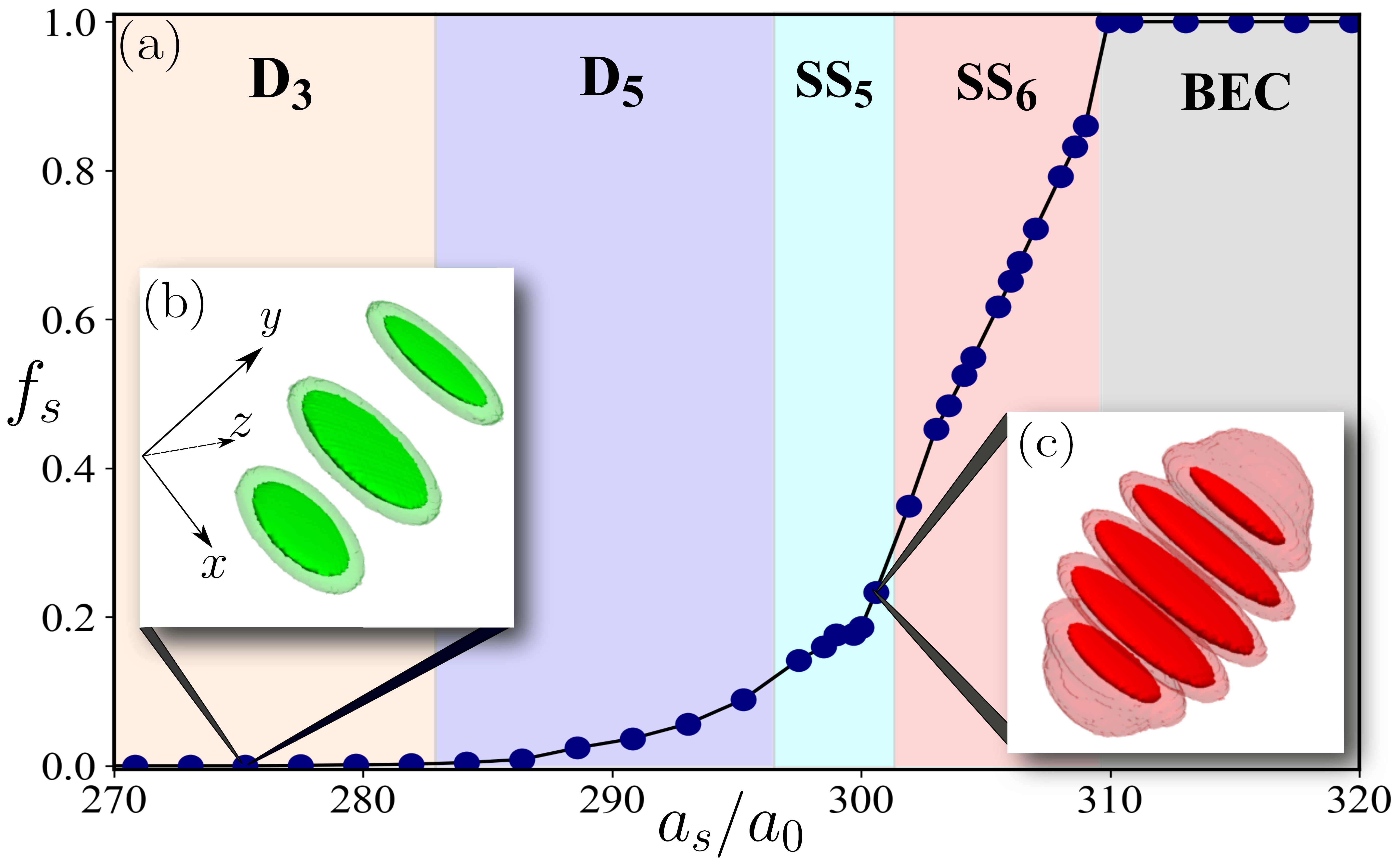}
  \caption{(Color online) Superfluid fraction $f_s$ as a function of $a_s$ for $\alpha/\pi=0.5$. The other parameters are the same as in Fig.~\ref{fig:10}. Insets (b) and (c) illustrate, respectively, the case of an incoherent and supersolid pancake droplet arrays.} 
    \label{fig:11}
\end{figure}

We consider at this point the same trap as above, but exchanging the $x$ and $y$ axes, $\omega_{x,y,z}=2\pi \times (53, 18.5, 81)$ Hz. 
As shown in Fig.~\ref{fig:10}, for $\alpha\lesssim 0.12\pi$, the phase diagram remains the same as in the previous case since the electric dipole moment is small, and the trap frequency along $z$~(the magnetic dipole direction) is unchanged. In contrast, for larger $\alpha$, the phase diagram drastically departs from that of Fig.~\ref{fig:6}. 
The transitions move to larger values of $a_s$, but their nature 
remains basically unaltered. Most remarkably, droplets become pancake-shaped on the $xz$-plane for $\alpha>0.4\pi$. As a result, doubly-dipolar condensates offer the unique possibility of realizing arrays of pancake droplets, as illustrated in Fig.~\ref{fig:11} for the case of $\alpha/\pi=0.5$. 
Upon decreasing $a_s$, the unmodulated BEC phase undergoes a transition to a pancake supersolid~(${\rm SS}_6$), followed by a pancake supersolid-supersolid (${\rm SS}_6-{\rm SS}_5$) transition. Eventually, it becomes an incoherent array of pancake droplets (${\rm D_5}$), and a further decrease in $a_s$ leads to arrays with lesser droplets. 

\section{summary}
\label{summ}
The interplay between doubly-dipolar interactions and harmonic confinement leads to novel possibilities for quantum droplet arrays. 
The relative orientation between the electric and magnetic dipole moments constitutes a novel control parameter that may be employed 
to drive intriguing scenarios, such as modulated droplets and supersolid-supersolid transitions. Moreover, changing the relative angle allows, without changing the external confinement, to transition from an array of cigar-shaped droplets, as those of usual dipolar condensates, to a novel array of pancake droplets. Although we have focused on one-dimensional droplet arrays, two-dimensional arrangements open new possibilities for other forms of supersolids and density patterns, as explored in usual dipolar condensates~\cite{zha19, her21, zha21, pol21, nor21, bla22}. The fascinating physics of two-dimensional quantum-stabilized doubly-dipolar condensates will be the subject of future studies.


\section{Acknowledgments}
We thank National Supercomputing Mission (NSM) for providing computing resources of 'PARAM Brahma' at IISER Pune, which is implemented by C-DAC and supported by the Ministry of Electronics and Information Technology (MeitY) and Department of Science and Technology (DST), Government of India.  R. N. further acknowledges DST-SERB for Swarnajayanti fellowship File No. SB/SJF/2020-21/19 and National Mission on Interdisciplinary Cyber-Physical Systems (NM-ICPS) of the Department of Science and Technology, Govt. Of India through the I-HUB Quantum Technology Foundation, Pune INDIA. C. M. thanks Early Career Fellowship by IIT Gandhinagar. L. S. acknowledges the Deutsche Forschungsgemeinschaft (DFG, German Research Foundation) -- under FOR2247, and under Germany's Excellence Strategy -- EXC-2123 Quantum-Frontiers -- 390837967.
\bibliographystyle{apsrev4-1}
\bibliography{libdbec.bib}
\end{document}